\documentclass[aps,prl,epsfig,twocolumn,showpacs,preprintnumbers,amsmath,amssymb,superscriptaddress,floatfix]{revtex4}
\usepackage{epsfig} \usepackage{dcolumn} \usepackage{bm}
 \def\CuII{Cu$^{2+}$ }     \def\Neel{N\'{e}el }

\begin{document}

\title{Scale-free antiferromagnetic fluctuations in the $s=1/2$
  kagome antiferromagnet herbertsmithite.}  
\author{M.~A. de Vries} \email{m.a.devries@physics.org}
\affiliation{School of Physics \& Astronomy,
  University of St-Andrews, the North Haugh, St Andrews, KY16 9SS, UK}
\affiliation{Laboratory for Quantum Magnetism, \'{E}cole Polytechnique F\'{e}d\'{e}rale de Lausanne (EPFL), Switzerland}
\affiliation{CSEC and School of Chemistry,
  The University of Edinburgh, Edinburgh, EH9 3JZ, UK}
\author{J.~R. Stewart} 
\affiliation{Institut Laue-Langevin, 6 rue
  Jules Horowitz, F-38042 Grenoble, France}
\affiliation{ISIS facility, Rutherford Appleton Laboratories, STFC,
  Chilton, Didcot OX11 0DE, UK}  
\author{P.~P. Deen}
\affiliation{Institut Laue-Langevin, 6 rue Jules Horowitz, F-38042
  Grenoble, France}
\author{J.~O. Piatek}
\affiliation{Laboratory for Quantum Magnetism, \'{E}cole Polytechnique
  F\'{e}d\'{e}rale de Lausanne (EPFL), Switzerland}
\author{G.~J. Nilsen}
\affiliation{Laboratory for Quantum Magnetism, \'{E}cole Polytechnique
  F\'{e}d\'{e}rale de Lausanne (EPFL), Switzerland}
\author{H.~M. R\o{}nnow} 
\affiliation{Laboratory for Quantum Magnetism, \'{E}cole Polytechnique
  F\'{e}d\'{e}rale de Lausanne (EPFL), Switzerland}
\author{A. Harrison}   
\affiliation{Institut Laue-Langevin, 6 rue
  Jules Horowitz, F-38042 Grenoble, France}  
\affiliation{CSEC and
  School of Chemistry, The University of Edinburgh, Edinburgh, EH9
  3JZ, UK}
\date{\today}


\begin{abstract} 
Neutron spectroscopy and diffuse neutron scattering on herbertsmithite
[ZnCu$_3$(OH)$_6$Cl$_2$], a near-ideal realisation of the $s=1/2$
kagome antiferromagnet, reveal the hallmark property of a quantum spin
liquid; instantaneous short-ranged  antiferromagnetic correlations in
the absence of a time-averaged ordered moment. These dynamic
antiferromagnetic correlations are weakly dependent of
neutron-energy transfer and temperature, and persist up to 25 meV and
120 K. At low energy transfers a shift of the magnetic scattering to
low $Q$ is observed with increasing temperature, providing evidence of
gapless spinons. It is argued that these observations provide
important evidence in favour of RVB theories of (doped) Mott
insulators.   
\end{abstract}


\maketitle{}

There has been a long search for quantum spin liquids in Mott
insulators in which the common \Neel antiferromagnetic order is
destabilised by quantum fluctuations~\cite{Anderson:87,Lhuillier:04}, the
mixing-in of spin-singlets.  The $s=1/2$ kagome antiferromagnet is one
system where such a spin liquid state could, in theory, be found. The
kagome lattice, which owes its name to a Japanese basket weaving
method, is a net of corner-sharing triangles. When
antiferromagnetically (AF)-coupled spins are arranged at the vertices
of this lattice it becomes impossible to satisfy all AF bonds
simultaneously. Due to this geometric frustration, the ground state
has a continuous and macroscopic degeneracy.

Both in the classical limit of large spin~\cite{Reimers:91, Ritchey:93}
and in the quantum limit of $s=1/2$~\cite{Chalker:92b}, the kagome
AF has been predicted to retain the full symmetry of the
magnetic Hamiltonian. Exact diagonalization (ED) results suggest that
in the macroscopic limit of the $s=1/2$ system a continuum of
non-magnetic global spin-singlet $s_{\text{tot}} = 0$ states fills an
energy gap to magnetic excitations~\cite{Misguich:07}.  Whether there
is a spin gap in real systems has however been debated and in most
theories building on Anderson's RVB proposal~\cite{Anderson:87}, such
as the algebraic spin liquid (ASL)~\cite{Affleck:88, Ran:07,
  Hermele:08}, there are gapless spinons~\cite{Affleck:88,
  Lee:RPP08}. Spinons are $s=1/2$ excitations which are created in
pairs following a singlet-triplet excitation but can dissociate with no
(additional) cost in energy. The ASL is thought of as a ``mother of
competing orders''~\cite{Hermele:05} including \Neel
antiferromagnetism and d-wave superconductivity. As the insulating analogue
for the Fermi liquid in metals the ASL would be a natural candidate
for the ground state in a Mott insulator where \Neel order is
suppressed. The results presented here on the kagome AF
compound herbertsmithite provide further evidence for this idea. At
the same time, the absence of a spin gap in herbertsmithite can be
reconciled with ED results~\cite{Lauchli:09} when a
Dzyaloshinky-Moriya interaction~\cite{Cepas:08b} or structural
disorder is taken into account. 

Of the many kagome systems studied experimentally, including the
jarosites~\cite{Grohol:05,Coomer:06}, SCGO
(SrCr$_{8-x}$Ga$_{4+x}$O$_{19}$)~\cite{Broholm:90} and
volborthite~\cite{Hiroi:01}, herbertsmithite
[ZnCu$_3$(OH)$_6$Cl$_2$]~\cite{Shores:05} is the first where no
(partial) freezing of the spins is observed down to the lowest
temperatures accessible experimentally~\cite{Mendels:07, Helton:07}.
The Weiss temperature is $300$~K and the AF exchange interaction $J
\approx 190$~K~\cite{Helton:07,Misguich:07}.  Herbertsmithite contains
well separated 2D kagome layers of \CuII ions linked by O$^{2-}$ ions
in OH$^-$ groups. Separating the kagome layers are Zn sites of $O_h$
symmetry, which can also host Cu$^{2+}$ ions to form the zinc
paratacamite family Zn$_x$Cu$_{4-x}$(OH)$_6$Cl$_2$ with $0 < x \leq
1$.  For $x=1$ the low-temperature susceptibility is dominated by a
Curie-like contribution from ``antisite spins'' (a $\sim 6$\% fraction
of Cu$^{2+}$ spins which have traded places with Zn$^{2+}$ to occupy
the interplane Zn site)~\cite{deVries:08, Bert:07}. These
weakly-coupled $s=1/2$ spins, which have been identified as individual
doublets~\cite{deVries:08}, have also been observed in neutron spectra
at the Zeeman energy in applied fields~\cite{Helton:07,
  SHLee:07}. Using  $^{17}$O NMR~\cite{Olariu:08} and Cl
NMR~\cite{Imai:07} there is now convincing evidence that the kagome
layers have a nonzero susceptibility as $T \to 0$. This is in
agreement with the continuum of magnetic excitations observed using
neutron spectroscopy at energies between $\sim 0.8$ and
2~meV~\cite{Helton:07}. Moreover, this spectrum is independent of the
temperature, which has been suggested to point to the proximity of a
quantum critical point~\cite{Helton:07}. Here we present neutron
spectroscopy data up to energy transfers of $\sim 30$~meV and
temperatures up to 120~K, to measure the dynamic magnetic structure
factor and fully characterize the quantum spin dynamics in
herbertsmithite. 

Large single crystals of herbertsmithite do not presently
exist. Hence, 20~g of 98\% deuterated herbertsmithite powder was
synthesised following the method as described in
Ref.~\onlinecite{Shores:05}. This sample was characterized using
neutron diffraction, DC magnetic susceptibility, heat
capacity~\cite{deVries:08} and $\mu$SR
measurements~\cite{Mendels:07}. The measurements were carried out
using the polarised neutron spectrometers D7 and IN22, and the
time-of-flight (TOF) spectrometers IN4, ILL, France, and MARI, ISIS,
UK. At D7 the magnetic scattering was separated from the nuclear
and spin-incoherent scattered neutrons using $XYZ$ polarisation
analysis~\cite{Stewart:09}. No energy analysis was carried out on the
scattered neutrons, with an incident neutron energy of 8.95~meV,
effectively energy integrating the neutron cross
section~\cite{footnote1} up to energy transfers ($\epsilon$) of $\sim
6.5$~meV. Complete $S(Q,\epsilon)$ maps without polarisation analysis
were obtained at 2, 4, 10, 30, 60 and 120~K at IN4 using 17.21~meV
neutrons and at 2K at MARI using 56~meV neutrons. The normalised spin
and nuclear incoherent signals measured at D7 were used to normalise
the IN4 and MARI data. Neutron polarisation and energy resolved
measurements were carried out at the triple-axis spectrometer IN22,
ILL. Several $Q$, $\epsilon$ and $T$ scans were performed for spin
flip (SF) and non spin flip (NSF) channels for incident and scattered
neutron spins polarised parallel to $Q$ (labelled $xx$) and
perpendicular to $Q$ (labelled $zz$). 

\begin{figure}[htbp] 
\epsfig{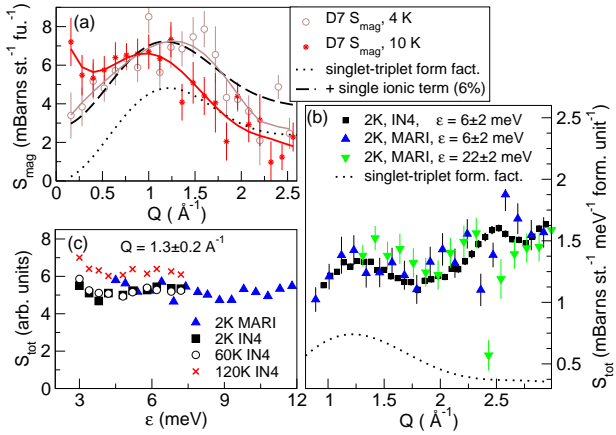}
\caption{(Color online) a) Instantaneous magnetic correlations at 4 K
  (open circles) and 10 K (filled circles) from D7. Data taken at 60~K
  (not shown) resembles the 10~K data. The solid lines are a guide to
  the eye.  b) The $Q$ dependence in the dynamic correlations from IN4
  and MARI with the energy integration interval indicated in the
  legend.  The dotted line in panel a and b is the structure factor
  for dimer-like AF correlations, for the dashed line a single-ion
  contribution corresponding to the 6\% antisite spins in this system
  is added. c) The energy and temperature dependence at $Q =
  1.3$~\AA$^{-1}$ from the TOF data.}
\label{figure:Qmag}
\end{figure}

At D7 the energy of the scattered neutrons is not analysed. The D7
data (Fig.~\ref{figure:Qmag}(a)) therefore correspond to magnetic
correlations which persist on a time scale of at most~\cite{footnote1}
$\sim$ 6.5 meV (1.6 THz). The broad peak observed here at
1.3~\AA$^{-1}$ is direct evidence of short-ranged instantaneous
near-neighbour AF correlations. For comparison the powder-averaged
structure factor for uncorrelated near-neighbour AF dimers given by
\begin{equation} 
F^2(Q)\left( 1- \frac{\sin(Qd)}{Qd} \right),
\label{equation:dimerFF}
\end{equation} 
is shown in Fig.~\ref{figure:Qmag}(a) and (b). In
Eq.~\ref{equation:dimerFF} $F(Q)$ is the \CuII magnetic form factor
and $d=a/2=3.42$~\AA~the Cu-Cu distance within the kagome plane. 

\begin{figure}[htbp] 
\epsfig{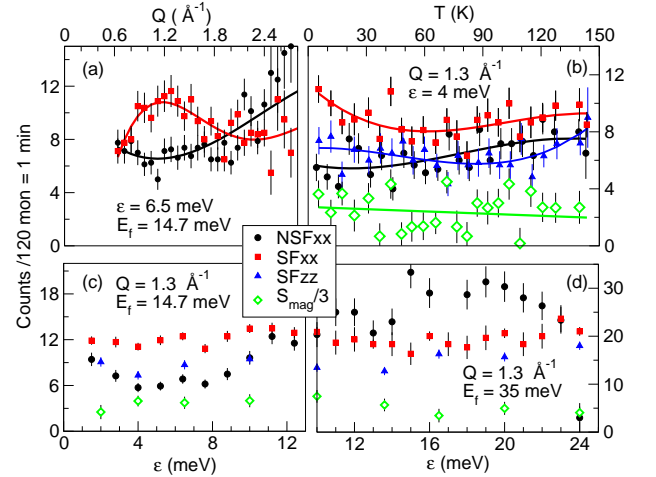}
\caption{(Color online) Inelastic intensity for 3 polarisation channels: NSF$_{xx}$
  contains phonons and NSF background. SF$_{xx}$ contains 2/3 of the
  magnetic intensity and SF background. SF$_{zz}$ contains 1/3 of the
  magnetic intensity. Hence, $S_{\text{mag}}/3 = $ SF$_{xx}-$ SF$_{zz}$.
  It is seen that magnetic intensity at the maximum at $Q =
  1.3$~\AA$^{-1}$ (SF$_{xx}$ in panel a) is finite and nearly constant
  for all covered temperatures (b) and energies (c,d). Data in
  (a),(c) and (d) were measured at 2K.
}
\label{figure:IN22}
\end{figure}

Similar short-ranged AF correlations are observed in the TOF data from IN4 and MARI
at all  $\epsilon$ accessed, up to 25~meV. Fig.~\ref{figure:Qmag}(b)
and (c) show respectively the momentum and energy dependence of the
magnetic scattering peak in the TOF data. Fig.~\ref{figure:IN22} shows
scans in $\epsilon$, temperature and $Q$ using polarisation analysis
which confirm that these dynamic correlations are of magnetic origin.
That a similar $Q$ dependence is observed in the D7 data suggests that
the dynamic magnetic correlations remain unchanged down to the elastic
line.   From Fig.~\ref{figure:IN22}(b,c,d) it is clear that in
addition to magnetic inelastic scattering ($S_{\text{mag}}$) there is
significant inelastic background from spin incoherent
(SF$_{\text{inc}} = $ SF$_{zz}-S_{\text{mag}}$) and nuclear
scattering. This is ascribed to  incoherent scattering from residual
protons and multiple scattering. This background is also present in
the IN4 data and to a lesser extent in the MARI data because there a
smaller sample was mounted in an annular geometry. The integrated
intensity of the dynamic AF correlations as found in
Fig.~\ref{figure:Qmag} and \ref{figure:IN22} up to 25 meV corresponds
to 30 to 50\% of the sum rule $s(s+1)$ for the spins on the kagome
lattice.

In the raw IN4 data the dynamic
correlations at $Q<1.8$~\AA$^{-1}$ persist up to 30~K while
the magnetic scattering at the maximum of the peak at $1.3$~\AA$^{-1}$
changes very little up to 120~K
(Fig.~\ref{figure:Qmag}(c) and Fig.~\ref{figure:IN22}(b)). 
The intensity at the energy loss side ($\epsilon <0$) in the TOF data
obeys detailed balance at all temperatures. This combined with the
magnetic scattering intensity and temperature dependence in the D7
data implies that at the elastic line there must be increased
magnetic scattering, but still with the same $Q$ dependence. These
(quasi) static magnetic correlations are reduced with a shift of
intensity to lower $Q$ as the temperature is increased, in what looks
like spinon excitations in the D7 data at 10~K
(Fig.~\ref{figure:Qmag}(a)).  

The above observations complement previous neutron spectroscopy
results on herbertsmithite~\cite{Helton:07}, which show the inelastic
magnetic scattering cross section is energy independent between 0.8
and 2~meV apart from a weak field-dependent peak which is due to the
weakly-coupled Cu spins on Zn sites. The $Q$ dependence in this energy
range is not peaked at $1.3$~\AA$^{-1}$ which is due to the single ion
contribution from antisite spins. This contribution also needs to be
added to Eq.~\ref{equation:dimerFF} to fit the D7 data. The absence of
a peak around the edge of the first Brillouin zone in
Ref.~\onlinecite{Helton:07} down to 35 mK implies there is no
significant increase of the dynamic magnetic correlations over three
orders of magnitude in temperature! 

\begin{figure}[htbp] 
\epsfig{file=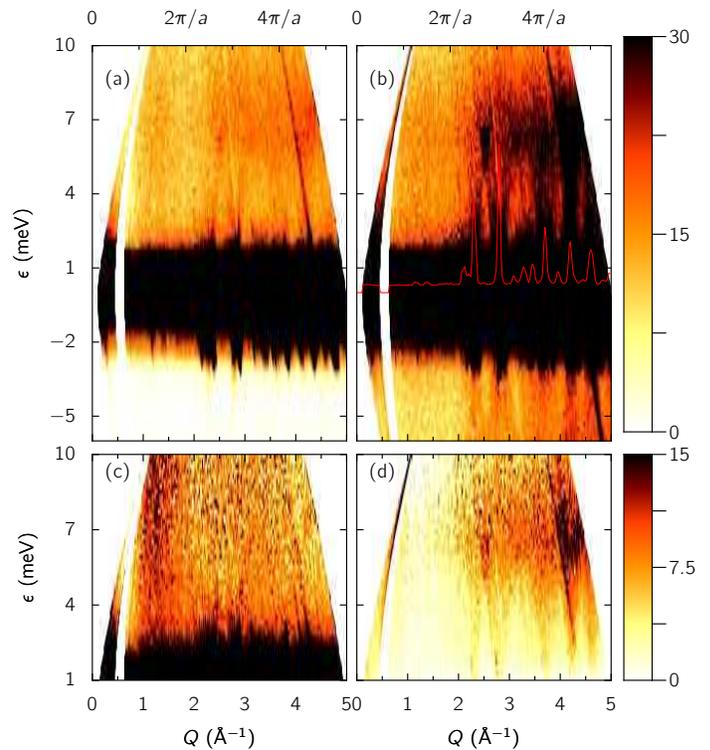, width = 3.6 in}
\caption{(Color online) The scattering cross section $S(Q,\epsilon )$
  of at 2~K (a) and 120 K (b). The intensity scale is mBarn st.$^{-1}$
  meV$^{-1}$ form. unit$^{-1}$. The elastic powder diffraction pattern
  is also shown in (b). Panel (c) and (d) show respectively
  $S_{\text{AF}}$ and $\chi''$. The white band at low $Q$ is a gap
  between detectors. Some increase in intensity in $S_{\text{AF}}$ is
  observed at higher energies but this is likely due to the direct beam
  and larger error bars. As shown in Fig.~\ref{figure:cutS}(a) the
  increase in intensity is only very small.} 
\label{figure:IN4}
\end{figure}

\begin{figure}[htbp] \epsfig{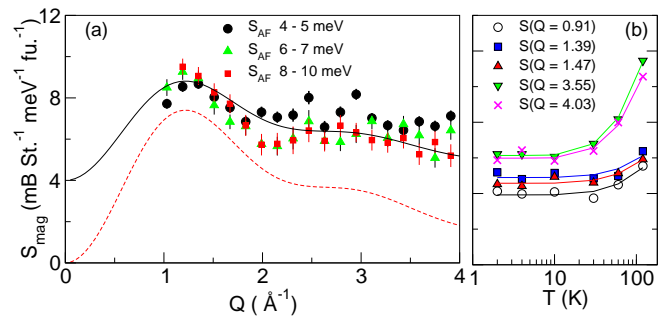}
\caption{(Color online) a) The $Q$ dependence in $S_{\text{AF}}(Q,\epsilon)$ for a
  number of energy transfers. b) The temperature
  dependence for a number of points in $S(Q,\epsilon=4\text{~meV})$
  and fits to the data using Eq.~\ref{equation:fitf}.}
\label{figure:cutS}
\end{figure}

At $Q >1.8$~\AA$^{-1}$ and for $T>30$~K the inelastic scattering is
increasingly dominated by phonons. The temperature dependence of
phonon scattering can in general be calculated from a temperature
independent dynamic susceptibility using the fluctuation dissipation
theorem. The temperature independence of the magnetic scattering
observed at low $Q$ prompted us to fit the temperature dependence of
each pixel in $S(Q,\epsilon; T)$ (Fig.~\ref{figure:IN4}(a) and (b) for
$T = 2$~K and 120 K respectively) as the sum of a temperature
independent component $S_{\text{AF}}(Q,\epsilon)$ and a component
following linear  response $\chi ''(Q,\epsilon)$,  
\begin{equation}  
S(Q,\epsilon;T)=S_{\text{AF}}(Q,\epsilon) + (1-{\rm e}^{-\epsilon
  /k_{\rm B}T})^{-1} \chi''(Q,\epsilon). 
\label{equation:fitf}
\end{equation}
The $S_{\text{AF}}(Q,\epsilon)$ and $\chi ''(Q,\epsilon)$ resulting
from the fit are shown in Fig.~\ref{figure:IN4}(c) and (d)
respectively.  As expected, $S_{\text{AF}}(Q,\epsilon)$ corresponds to
the raw data from IN4 at low $Q$ and $T< 60$~K and there is a good
overall agreement with the structure factor of
Eq.~\ref{equation:dimerFF} (Fig.~\ref{figure:cutS}(a)) added to a
constant background.  Phonon dispersion originating from the nuclear
Bragg peaks is on the other hand clearly visible in $\chi
''(Q,\epsilon)$. The excellent fit for all data at 6 different
temperatures, as illustrated in Fig.~\ref{figure:cutS}(b) for a small
sample of points, confirms that the magnetic correlation length does
not diverge as the temperature is lowered and that these dynamic
correlations persist up to at least 120~K.  The maximum in $\chi''$
and $S(Q,\epsilon)$ around 7~meV that extends to $Q<1.8$~\AA$^{-1}$
gives the impression of an increase in magnetic scattering at
$(Q,\epsilon) = (1.3$~\AA$^{-1} ,7$~meV$)$. This signal in $\chi''$ at
low $Q$ is most likely due to some weak multiple scattering at IN4
because it was not observed in the MARI and IN22 data.

Hence, we found that 1) At low frequencies ($\epsilon < 2$~meV) there is
a shift of intensity to lower $Q$ as the temperature increases, which
could be due to (gapless) spinons. Spinons then also account for the
non-zero magnetic susceptibility as measured with $^{17}$O
NMR~\cite{Olariu:08}. 2) There are dynamic AF correlations with a very
weak temperature and energy dependence, extending over more than three
orders of magnitude in temperature and two orders of magnitude in
energy. Hence, the dynamic AF correlations are approximately
scale free. There is no characteristic energy scale in the system, not
even one determined by temperature as in systems with $\omega/T$
scaling (with $\omega=\epsilon/\hbar$) and as expected for the
ASL~\cite{Hermele:08}. Although with neutron scattering any long range order of
spin-singlet dimers cannot be detected directly, the absence of a spin
gap and the temperature independent dynamic correlations observed here
and in~\cite{Helton:07} imply the absence of a static dimerisation, at
least down to 35 mK. Short ranged dynamic correlations are commonly
observed in quantum spin liquids at temperatures well above their
respective ordering and spin-glass transitions~\cite{Shirane:87,
  Broholm:90, Schweika:07}. The temperature independence of the
dynamic AF correlations observed here suggests that, intriguingly, the
short-ranged dynamic correlations could be a property of the ground
state in herbertsmithite.

A system of particular relevance in the present context  is the
spin-glass phase in La$_{2-x}$Sr$_x$CuO$_4$ (LSCO) with doping around
$x=0.05$~\cite{Keimer:91, Hayden:91}. At this level of doping, just
enough for the N\'eel order to be completely suppressed, the
correlation length of the dynamic AF correlations becomes temperature
independent not far below the N\'eel temperature of the parent
compound. Such a strong likeness with the dynamic magnetic
correlations in herbertsmithite is in agreement with RVB theories,
which predict that the Mott insulator in the absence of N\'eel order
and the pseudogap phases are closely related states.  One
proposal is that both are described by a staggered-flux, or algebraic
spin liquid~\cite{Affleck:88, Hermele:08}, a state with algebraically
diverging correlations. Here the agreement with
the spatial and temporal correlations in 
herbertsmithite, and with the spatial correlations in LSCO, breaks
down. This could be due to weak structural disorder in the
respective Cu-O planes which both systems are known to suffer from. 
To test this possibility new
herbertsmithite samples are needed with reduced antisite disorder. The
energy spectrum and powder-averaged $Q$ dependence measured here are,
apart from the absence of a spin gap, also in rough agreement with ED
results~\cite{Lauchli:09} but large single crystals and theoretical
predictions on the temperature dependence would be needed for a more
detailed comparison.

L.-P.~Regnault (CEA, Grenoble) is acknowledged for assistance with the
IN22 measurements. We are also grateful for fruitful discussions with
D.~Ivanov and S.~Bieri (EPFL), O. Cepas (Institut N\'eel, Grenoble),
Jorge Quintanilla (ISIS, RAL), A.~Huxley (Univeristy of Edinburgh),
J.~Zaanen (Leiden University) and J. Chalker (Oxford University). MdV
acknowledges an exchange grant from the HFM network of the European
Science Foundation.


\begin{thebibliography}{10}
\bibitem{footnote1}
The upper limit for the frequencies which are probed are a function of
$Q$, approaching 0 as $Q\to 0$.
\end{thebibliography}
\end{document}